\documentclass[twoside,twocolumn,english,pra,showpacs]{revtex4}
\usepackage[T1]{fontenc}
\usepackage[latin9]{inputenc}
\usepackage{amsmath}
\usepackage{graphicx}
\usepackage{amssymb}
\usepackage{esint}

\makeatletter
\@ifundefined{textcolor}{}
{%
 \definecolor{BLACK}{gray}{0}
 \definecolor{WHITE}{gray}{1}
 \definecolor{RED}{rgb}{1,0,0}
 \definecolor{GREEN}{rgb}{0,1,0}
 \definecolor{BLUE}{rgb}{0,0,1}
 \definecolor{CYAN}{cmyk}{1,0,0,0}
 \definecolor{MAGENTA}{cmyk}{0,1,0,0}
 \definecolor{YELLOW}{cmyk}{0,0,1,0}
 }

\usepackage{array}\usepackage{amsbsy}\usepackage{amstext}

\makeatletter

\@ifundefined{textcolor}{}
{%
 \definecolor{BLACK}{gray}{0}
 \definecolor{WHITE}{gray}{1}
 \definecolor{RED}{rgb}{1,0,0}
 \definecolor{GREEN}{rgb}{0,1,0}
 \definecolor{BLUE}{rgb}{0,0,1}
 \definecolor{CYAN}{cmyk}{1,0,0,0}
 \definecolor{MAGENTA}{cmyk}{0,1,0,0}
 \definecolor{YELLOW}{cmyk}{0,0,1,0}
 }

\makeatother

\usepackage{babel}

\makeatother

\usepackage{babel}

\begin{document}

\title{Synthetic magnetic field effects on neutral bosonic condensates \\in
quasi three-dimensional anisotropic layered structures}

\author{T. A. Zaleski}

\affiliation{Institute of Low Temperatures and Structure Research, Polish Academy
of Sciences, POB 1410, 50-950 Wroc\l{}aw 2, Poland}

\author{T. P. Polak}

\affiliation{Adam Mickiewicz University of Pozna\'{n}, Faculty of Physics, Umultowska
85, 61-614 Pozna\'{n}, Poland}
\begin{abstract}
We discuss a system of dilute Bose gas confined in a layered structure
of stacked square lattices (slab geometry). A derived phase diagram
reveals a non-monotonic dependence of the ratio of tunneling to on-site
repulsion on the artificial magnetic field applied to the system.
The effect is reduced when more layers are added, which mimics a two-
to quasi-three-dimensional geometry crossover. Furthermore, we establish
a correspondence between anisotropic infinite (quasi three-dimensional)
and isotropic finite (slab geometry) systems that share exactly the
same critical values, which can be an important clue for choosing
experimental setups that are less demanding, but still leading to
the identical results. Finally, we show that the properties of the
ideal Bose gas in a three-dimensional optical lattice can be closely
mimicked by finite (slab) systems, when the number of two-dimensional
layers is larger than ten for isotropic interactions or even less,
when the layers are weakly coupled. 
\end{abstract}


\maketitle

\section{Introduction}

Systems of dilute bosonic gases confined in optical lattices are ideal
toolboxes for testing theoretical models and their solutions \cite{jaksch_toolbox}.
However, investigation of their properties under external magnetic
field seemed to be precluded, since the particles used in the experiments
are uncharged atoms and thus, are not directly affected by the magnetic
field. Fortunatelly, development of various experimental techniques
allow to investigate ultracold systems that are described by exactly
the same Hamiltonians as the ones interacting with external magnetic
field. Furthermore, they lead to appearance of vortices in Bose-Einstein
condensates (BEC), which is a hallmark of a superfluid in a magnetic
field. One of those techniques results from equivalence between the
Lorentz force and the Coriolis force. As a result, rotation of the
condensate (usually rotation of the masks with set of holes located
in the laser beams producing quasi-two-dimensional rotating optical
lattice) acts as a synthetic magnetic field (SMF) \cite{coddington,tung,schweikhard}.
However, this approach puts limit on maximum rotational velocity,
thus large SMF (e.g. required for quantum Hall physics) cannot be
reached. To overcome those difficulties, imprinting of the quantum
mechanical phase is used, which is based on superimposing of an external
potential on a BEC (e.g. by applying rotating magnetic field) \cite{leanhardt,zheng}.
Another approaches are stirring a BEC with laser beam \cite{madison},
or using additional Raman lasers to coherently transfer atoms from
one internal state to another. This induces a non-vanishing phase
of particles moving along a closed path, which simulates magnetic
flux through the lattice \cite{jaksch,lin}. Additionally, rectification
of the magnetic field in the optical lattice by using a superlattice
allows to ensure that each plaquette acquires the same phase, thus
simulating the uniform magnetic field for any value of phase between
$0$ and $\pi$ \cite{gerbier}.

A ground state of neutral atoms in an optical trapping potential can
be either superfluid (SF) or a Mott-insulator (MI). The zero-temperature
coupling ($t/U$) vs. chemical potential ($\mu/U$) phase diagram
contains characteristic lobes marking a quantum phase transition between
SF and MI states; $t$ is the matrix element for tunneling between
adjacent lattice sites and $U$ is the on-site energy cost for multiple
occupancy. For average number of particles per site equal to one,
the transition occurs at the ratio $\left(t/U\right)_{\mathrm{crit}}$,
which is strongly dependent on the geometry of the system. Recently,
many very precise calculations and measurements have been performed
to obtain correct value of the $\left(t/U\right)_{\mathrm{crit}}$.
For the three-dimensional ($3D$) optical lattices, most methods converge
to $\left(t/U\right)_{\mathrm{crit}}^{3D}=0.03$ \cite{capogrosso-sansone,polak},
but lower-dimensional cases are more demanding because of growing
influence of the quantum fluctuations. In order to precisely describe
experimental results, it is necessary to develop a theory that contains
fully tunable lattice degrees of freedom, includes effects of the
synthetic magnetic field and is nonperturbative, which allow to capture
essential physics of strongly correlated system in $U/t\gg1$ regime.
To this end, we constructed a theoretical field approach, in which
the dimensionality along $c$ axis can be tuned by adding an arbitrary
number of single layers along with variable tunneling between them
(thus including anisotropy of tunnelling between and within planes).
In principle, we can include sixty layers as in the experiments of
the Spielman's group \cite{spielman}, or just a few as in the Krüger
setup \cite{kruger}, where the magnetic confinement along the $z$-direction
was used to localize atoms in the effectively $\mathcal{N}=4k_{\mathrm{B}}T/m\omega_{z}^{2}l^{2}\sim2\div4$
central lattice planes ($T$ is the temperature, $\omega_{z}$ is
the frequency at the bottom of the lattice wells and $l$ is the lattice
period that can be adjusted to any value higher than half of the atomic
resonance wavelength). The Burger group \cite{burger} developed method
in which by increasing the nodal planes of the optical lattice superposed
to a $3D$ potential it is possible to follow the transition from
three-dimensional system to an two-dimensional array. Therefore the
system becomes quasi-$3D$ rather than purely planar. That can be
used in order to look in the interference between planes and in consequence
to access to spatial coherence of such structure.

The outline of the paper is as follows: in Sec. II we introduce the
model Hamiltonian and the effects of artificial magnetic field and
briefly describe our method. In Sec. IV, we present our results starting
with the zero-temperature phase diagrams and its dependence on the
system geometry. Furthermore, we show the influence of the synthetic
magnetic field on properties of the Bose-Hubbard system. Also, we
find a correspondence between systems sharing the same critical properties
but differing with the geometry of the opticall lattice and interactions,
which can be a helpful clue on choosing between various experimental
setups. Finally, we conclude in Sec. V.

\section{Model}

In optical lattices, two main energy scales are set by the hopping
amplitude $t$ (the kinetic energy of bosons tunneling between the
lattice sites), and the on-site repulsive interaction $U$ (resulting
from repulsion of multiple boson occupying the same lattice site).
For $t\gg U$, the superfluid order is well established in zero-temperature
limit. However, for sufficiently large repulsive energy $U$, the
quantum phase fluctuations lead to suppression of the long-range phase
coherence resulting in SF to MI transition. The critical ratio of
$\left(t/U\right)_{c}$ for which this transition occurs depends strongly
on the number of bosons introduced to the optical lattice (which in
theoretical models is often controled by a chemical potential $\mu$).
The synthetic magnetic field $\mathbf{B}$ (resulting either from
rotation of the system, phase imprinting, or external electric field)
introduces the Peierls phase factor $e^{\frac{2\pi i}{\Phi_{0}}\int_{\boldsymbol{r}_{j}}^{\boldsymbol{r}_{i}}\boldsymbol{\mathrm{A}}\cdot d\boldsymbol{l}}$,
where $\boldsymbol{B}=\nabla\times\boldsymbol{\mathrm{A}}\left(\boldsymbol{r}\right)$,
and $\Phi_{0}=hc/e$ is the flux quantum, with $\mathbf{A}\left(\mathbf{r}\right)$
being the vector potential (which can be realized experimentally,
see Ref. \cite{lin}), and $h$, $c$ and $e$ -- Planck constant,
speed of light and charge of electron, respectively. Thus, the system
can be described by the following quantum Bose-Hubbard Hamiltonian
\cite{fisher,polak1}\begin{eqnarray}
\mathcal{H} & = & \frac{U}{2}\sum_{\boldsymbol{r}}n_{\boldsymbol{r}}\left(n_{\boldsymbol{r}}-1\right)\nonumber \\
 & - & \sum_{\left\langle \boldsymbol{r,}\boldsymbol{r'}\right\rangle }t_{\boldsymbol{rr'}}e^{\frac{2\pi i}{\Phi_{0}}\int_{\boldsymbol{r}_{j}}^{\boldsymbol{r}_{i}}\boldsymbol{\mathrm{A}}\cdot d\boldsymbol{l}}a_{\boldsymbol{r}}^{\dagger}a_{\boldsymbol{r'}}-\mu\sum_{\boldsymbol{r}}n_{\boldsymbol{r}},\label{hamiltonian}\end{eqnarray}
where $a_{\boldsymbol{r}}^{\dagger}$ and $a_{\boldsymbol{r'}}$ are
for the bosonic creation and annihilation operators that obey canonical
commutation relations $[a_{\boldsymbol{r}},a_{\boldsymbol{r'}}^{\dagger}]=\delta_{\boldsymbol{r}\boldsymbol{r}'}$,
$n_{\boldsymbol{r}}=a_{\boldsymbol{r}}^{\dagger}a_{\boldsymbol{r}}$
is the boson number operator on the site $\boldsymbol{r}$. Here,
$\left\langle \boldsymbol{r},\boldsymbol{r'}\right\rangle $ denotes
summation over the nearest-neighbor sites. Furthermore, $t_{\boldsymbol{r}\boldsymbol{r}'}$
is the hopping matrix element with the dispersion $t_{\mathbf{k}}=2t_{\parallel}\left(\cos k_{x}+\cos k_{y}+\frac{t_{\perp}}{t_{\parallel}}\cos k_{z}\right)$
where $t_{\perp}$ is the hopping between layers and $t_{\parallel}$
within the planes. Since, we are interested in investigating the influence
of the lattice geometry on  the system properties, we consider a stack
of an arbitrary number ($L$) of two-dimensional planes coupled with
$t_{\perp}$. As a result, the values of $k_{x}$ and $k_{y}$ are
continuous ($k_{x,y}=-\pi,\dots,\pi$), while $k_{z}$ is descrete
($k_{z}=\frac{2\pi}{L}l$, where $l=0,\dots,L-1$). Also, we allow
for $c$-axis anisotropy, which is a ratio of inter-plane to in-plane
hopping $\eta=t_{\perp}/t_{\|}$. 

A boson hopping around a lattice cell of the area of $A$ will gain
an additional phase $2\pi f$ resulting from the synthetic magnetic
field, where $f=ABe/2\pi\hbar$. As a result, the properties of the
system will be periodic with a period corresponding to $1/f$. As
a result, the periodic potential leads to splitting of Landau levels
into integer number $q$ of sub-bands. Of special interest are the
values of the SMF which correspond to rational numbers of $f\equiv p/q=1/2,1/3,1/4,...$
($p$ is an integer), since for those values the energy spectra and
the density of states can be obtained exactly, although it is analytically
feasible only for small values of $q$. In the present paper, we present
new results for $f=1/8$ and $f=3/8$, which up to now have been analitically
unaccessible (see, Ref. \cite{polak-future}). Since, all properties
of the Hamiltonian Eq. (\ref{hamiltonian}) are invariant under $f\rightarrow-f$
and also under $f\rightarrow f+1$, it is sufficient to consider $f$
in the range $0<f<1/2$. In solid state physics obtaining $f=1/2$
in experimental setup would require magnetic field of the order of
$10^{5}T$. In optical lattices however, due to larger lattice spacings
(and larger $A$ consequently), much smaller values of \textbf{$\mathbf{B}$}
of the order of $10^{-3}T$ are required. This makes investigation
of the above-mentioned rational values of $f$ reasonable. 

To proceed, we rely on the quantum rotors approach. The method is
extensively described in Refs. \cite{polak,polak1,polak-1,kopec},
so here we only summarize its main points. We use the functional integral
representation of the model with bosonic operators becoming complex
fields $a_{\mathbf{r}}\left(\tau\right)$ (where $\tau$ is imaginary
Matsubara's time). The most important element of our method is a local
gauge transformation to the new bosonic variables:\begin{equation}
a_{i}\left(\tau\right)=b_{i}\left(\tau\right)\exp\left[i\phi_{i}\left(\tau\right)\right].\end{equation}
This allows to cast the strongly correlated bosonic problem into a
system of weakly interacting bosons, submerged into the bath of strongly
fl{}uctuating gauge potentials on the high energy scale set by $U$.
It also allows to formulate the problem in the phase representation,
which is the best suited to describe MI to SF transition, since it
is governed by phase fluctuations. As a result, the superfluid order
parameter can be written as:

\begin{equation}
\Psi_{B}\equiv\left\langle a_{i}\left(\tau\right)\right\rangle =b_{0}\psi_{B},\end{equation}
where non-zero value of $\psi_{B}$=$\left\langle \exp\left[i\phi_{i}\left(\tau\right)\right]\right\rangle $
results from phase ordering and $b_{0}$ is the amplitude of the bosonic
field:\begin{equation}
b_{0}^{2}=\left[4+2\left(1-\frac{1}{L}\right)\frac{t_{\perp}}{t_{\parallel}}\right]\frac{t_{\parallel}}{U}+\frac{\mu}{U}+\frac{1}{2}.\end{equation}
The coefficient $4+2\left(1-\frac{1}{L}\right)\frac{t_{\perp}}{t_{\parallel}}$
is an effective number of nearest neighbors averaged over all lattice
sites. In the zero-temperature limit we arrive at the equation for
the phase order parameter:\begin{equation}
1-\psi_{B}^{2}=\frac{1}{2N}\sum_{\mathbf{k}}\frac{1}{\sqrt{\frac{J_{\boldsymbol{k}=0}-J_{\boldsymbol{k}}}{U}+\upsilon^{2}\left(\frac{\mu}{U}\right)}}\label{eq:order-param}\end{equation}
 with\begin{equation}
\upsilon\left(\frac{\mu}{U}\right)=\mathrm{frac}\left(\frac{\mu}{U}\right)-\frac{1}{2},\end{equation}
where $\mathrm{frac}\left(x\right)=x-\left[x\right]$ is the fractional
part of the number and $\left[x\right]$ is the floor function which
gives the greatest integer less then or equal to $x$ and the phase
stiffness $J_{\mathbf{k}}=b_{0}^{2}t_{\mathbf{k}}$. 

Since the number of layers $L$ in the system is finite, the summation
in Eq. \ref{eq:order-param} runs over discrete values of $k_{z}$
and continuous values of $k_{x}$ and $k_{y}$. However, because density
of states of a single layer under SMF ($\rho_{f}$ ) is known, we
explicitly derive the density of stated of the whole stack of $L$
coupled planes (for calculation details, see Ref. \cite{zaleski}):\begin{equation}
\rho_{f}^{L}\left(\eta,\xi\right)=\frac{1}{L}\sum_{k_{z}}\rho_{f}\left(\xi-\eta\cos k_{z}\right).\end{equation}
As a result, the critical line equation ($\psi_{B}=0$) including
the effects of SMF and $c$-axis anisotropy reads:\begin{equation}
1=\frac{1}{2}\int_{-\infty}^{+\infty}d\xi\frac{\rho_{f}^{L}\left(\eta,\xi\right)}{\sqrt{2\left(\xi_{\mathrm{0}}-\xi\right)b_{0}^{2}\frac{t_{\parallel}}{U}+\upsilon^{2}\left(\frac{\mu}{U}\right)}},\label{critical line final}\end{equation}
with $\xi_{0}$ being the half-width of the band dispersion for selected
value of $f=p/q$.

\section{Results}

\subsection{Phase diagram}

\begin{figure}
\includegraphics[scale=0.75]{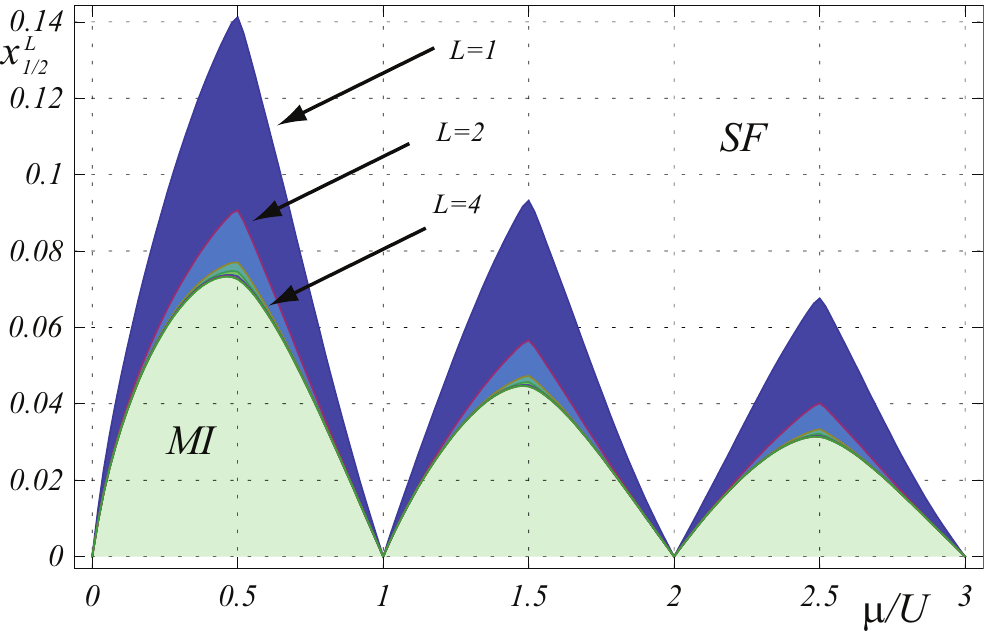}\caption{(Color online) The zero-temperature phase diagram of a stack of square
lattice planes (number of particles per lattice site is $n_{B}=1$
inside the first and $n_{B}=2$ and $3$ in second and third lobe,
respectively) with magnetic field $f=1/2$ for various number $L$
of layers and $\eta=1$. Within the MI phase the phase order parameter
$\Psi_{B}=0$.\label{phase diagram}}

\end{figure}

The Eq. (\ref{critical line final}) allows us to calculate the zero-temperature
phase diagram of the investigated Bose-Hubbard model from Eq. (\ref{hamiltonian})
as a dependence of critical interaction on the chemical potential,
SMF, number of layers and $c$-axis anisotropy:\begin{equation}
\left(\frac{t_{\|}}{U}\right)_{c}=x_{c}=x_{f}^{L}\left(\frac{\mu}{U},\eta\right).\label{eq:xc}\end{equation}
The diagram is plotted in Fig. \ref{phase diagram} for different
number of layers $L$, synthetic magnetic field $f=1/2$, in isotropic
case ($\eta=1$). In the weak coupling limit ($t_{\|}\gg U$), the
kinetic energy dominates and the ground state is a delocalized superfluid,
described by nonzero value of the superfluid order parameter $\Psi_{B}\neq0$.
On the other hand, in the strong coupling regime ($t_{\|}\ll U$)
the phase fluctuation become significant and the long-range order
is destroyed leading to a series of MI lobes with fixed integer filling
$n_{B}=1,2,...$\cite{polak,fisher}. A single-layer system ($L=1$)
has a simple square (two-dimensional) geommetry, which results in
the phase diagram with characteristic narrow-edged lobes. As the number
of layers is being increased, the tops of the lobes become smooth
and their maxima deviate towards lower values of the chemical potential
$\mu$ (this effect is also clearly presented in Ref. \cite{teichmann}).
As a result, the phase diagram becomes similar to the one of a cubic
(three-dimensional) system. It is important to notice, that also in
the presence of the synthetic magnetic field, the phase diagrams of
the finite $L$ system becomes indistinguishable from the infinite
(cubic) one for $L$ as small as $10$. For example, for ten layers,
the value of $x_{1/3}^{10}$ is $102.2\%$ of the infinite system
$x_{1/3}^{\infty}$ and for sixty -- $x_{1/3}^{60}$ is close to $100.1\%$
of $x_{1/3}^{\infty}$. The phase diagram obtained from the quantum
rotor approach (QRA) was verified several times \cite{polak1,polak-1}
by comparison with the very precise numerical methods: quantum Monte-Carlo
\cite{capogrosso-sansone} and diagrammatic perturbation theory \cite{teichmann}.
The QRA phase diagrams were also compared to an analytical works:
mean-field theory \cite{oktel} and Padé analysis \cite{niemeyer}.
From the above, it is expected that QRA accurately captures the low
energy physics of the Bose-Hubbard model in the more demanding case
where the system is placed under an artificial magnetic field.

\subsection{Effect of the synthetic magnetic field and the system geometry}

\begin{figure}
\includegraphics[scale=0.65]{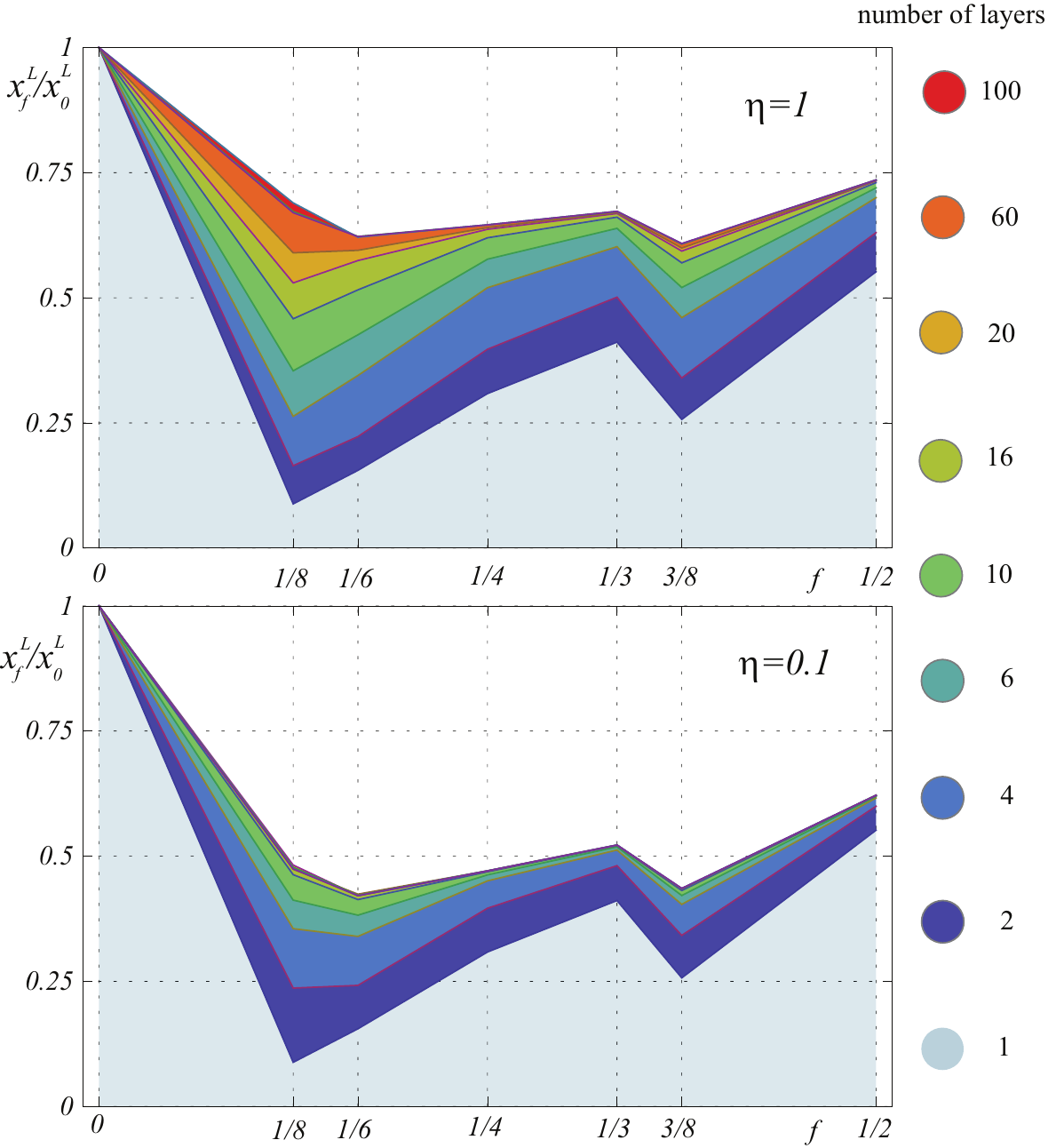}\caption{(Color online) Ratio of critical coupling $x_{0}^{L}/x_{f}^{L}$ of
the tip of the first lobe {[}see, Eq. (\ref{eq:xc}){]} of the system
without ($f=0$) and with synthetic magnetic field $f$ as a function
of number layers $L$ for isotropic (top) $\eta=1$ and anisotropic
(bottom) $\eta=0.1$ case, respectively.}

\label{nonmonotonical} 
\end{figure}
The transition, experimentally seen in the time-of-flight images (the
presence of sharp peaks has been considered as an unequivocal signature
of superfluidity in the Bose system), occurs rather rapidly with increasing
lattice depth. Because the experimental parameter $V_{0}/E_{R}$ ($V_{0}$
is the maximum value of the lattice depth), depends logarithmically
on $U/t$, the small changes of the dimensionless depth of the optical
lattice can cover a wide range of the phase diagram. The phase coherent
Bose gas can be also driven into the Mott insulating phase by applying
the synthetic magnetic field. The effect of the SMF is presented in
Fig. \ref{nonmonotonical}. The long range order is suppressed by
the phase changes imposed on the bosonic wave function and this suppression
has a non-monotonic character strongly depending on the topology of
the system. The Mott insulating phase becomes more stable, which is
as expected since the magnetic field should localize particles. In
the single-layer system, the effect of the SMF is the most pronunced
and this decreases with growing number of layers $L$. By adding more
layers the global coherence of the system is restored, because growing
dimensionality entails the suppression of quantum fluctuations effects.
Here, the convergence of properties of the finite system to those
of the infinite (cubic) one is much slower, although also non-trivially
dependent on $f$ (for some values like $f=1/8,1/6$ and $3/8$ seems
to be much more pronunced than for $f=1/4,1/3$ and $1/2$). When
the system is anisotropic (see, the bottom plot in Fig. \ref{nonmonotonical}),
the convergence is much faster, but still depending on the specific
value of $f$.

\begin{figure}
\includegraphics[scale=0.4]{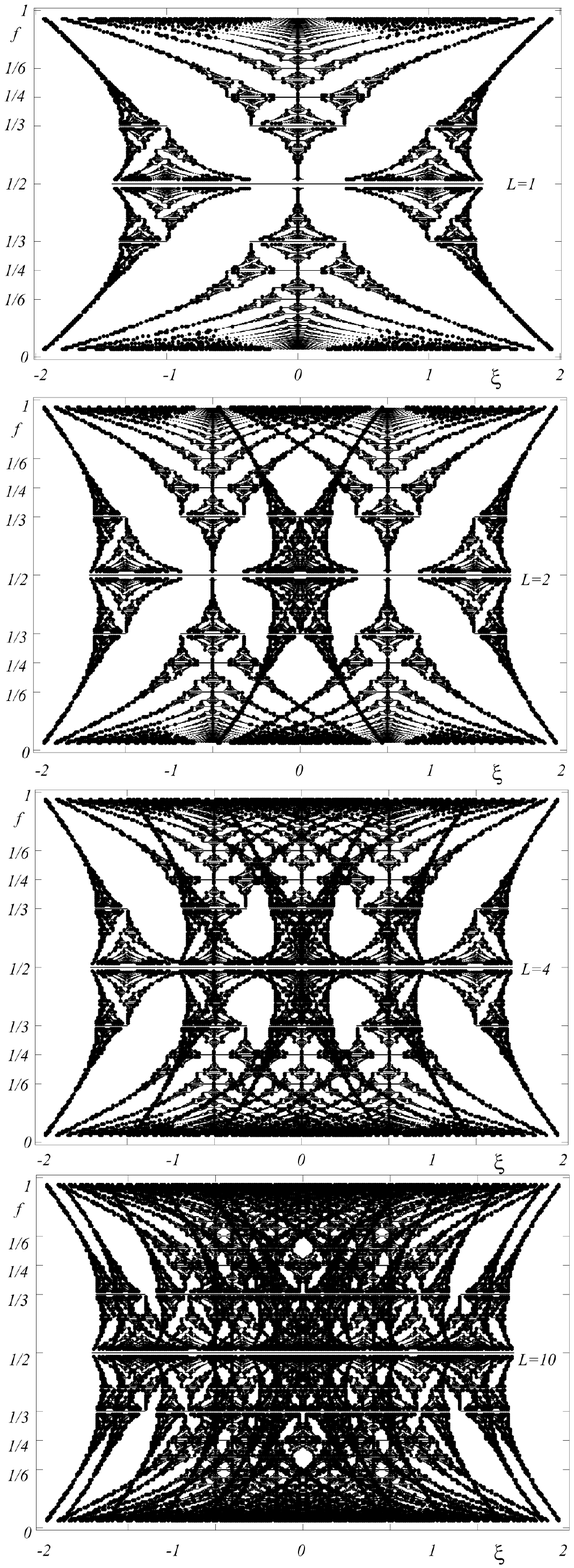}\caption{The evolution of the Hostadter butterfly with increasing number of
layers $L$ in the isotropic case of tunneling ratio $t_{\perp}=t_{\parallel}$.
Adding more layers simply results in additional fractal patterns and
very complex structures of $L$ butterflies emerging.}

\label{butterfly} 
\end{figure}

In the non-interacting ($U=0$) Bose-Hubbard model under the synthetic
magnetic field, the energy spectrum is known as the Hofstadter butterfly
\cite{hofstadter} (see, the top plot for $L=1$ in Fig. \ref{butterfly}).
The band width is strongly dependent on $f$ and exhibits a self-similar
gap structure for rational values of $f$. It has been observed, that
in the mean-field approach, the critical hopping $x_{f}^{1}$ roughly
follows the bandwidth of the Hofstadter's butterfly \cite{oktel}.
However, in our results the non-monotonicity of $x_{f}$ is not exactly
following the Hofstadter butterfly bandwidth, especially for lower
values of $f$. Also, we calculate the band for multiple-layer system
and present it in Fig. \ref{butterfly}. It is clear, that increasing
the number of layers (for $L>1$) does not influence the width of
the band, while the $x_{0}^{L}/x_{f}^{L}$ in Fig. \ref{nonmonotonical}
evidently changes while converging to the three-dimensional (cubic)
case. We would like to note that, both densities of states for square
system in the magnetic field $\rho_{f}$ and DOS of $L$-layer stack
$\rho_{f}^{L}\left(\eta,\xi\right)$ are obtained exactly in analytical
form with any approximations. The Hofstadter's band structure has
also been investigated in context of quantum graphs and the integer
quantum Hall effect\cite{goldman}, where similar energy spectra for
two-dimensional to three-dimensional system crossover have been observed
(non-symmetricity of the spectra results from their representation
as functions of wave vectors instead of energy).

\subsection{Correspondence between finite $L$-layer and infinite systems}

\begin{figure}
\includegraphics[scale=0.75]{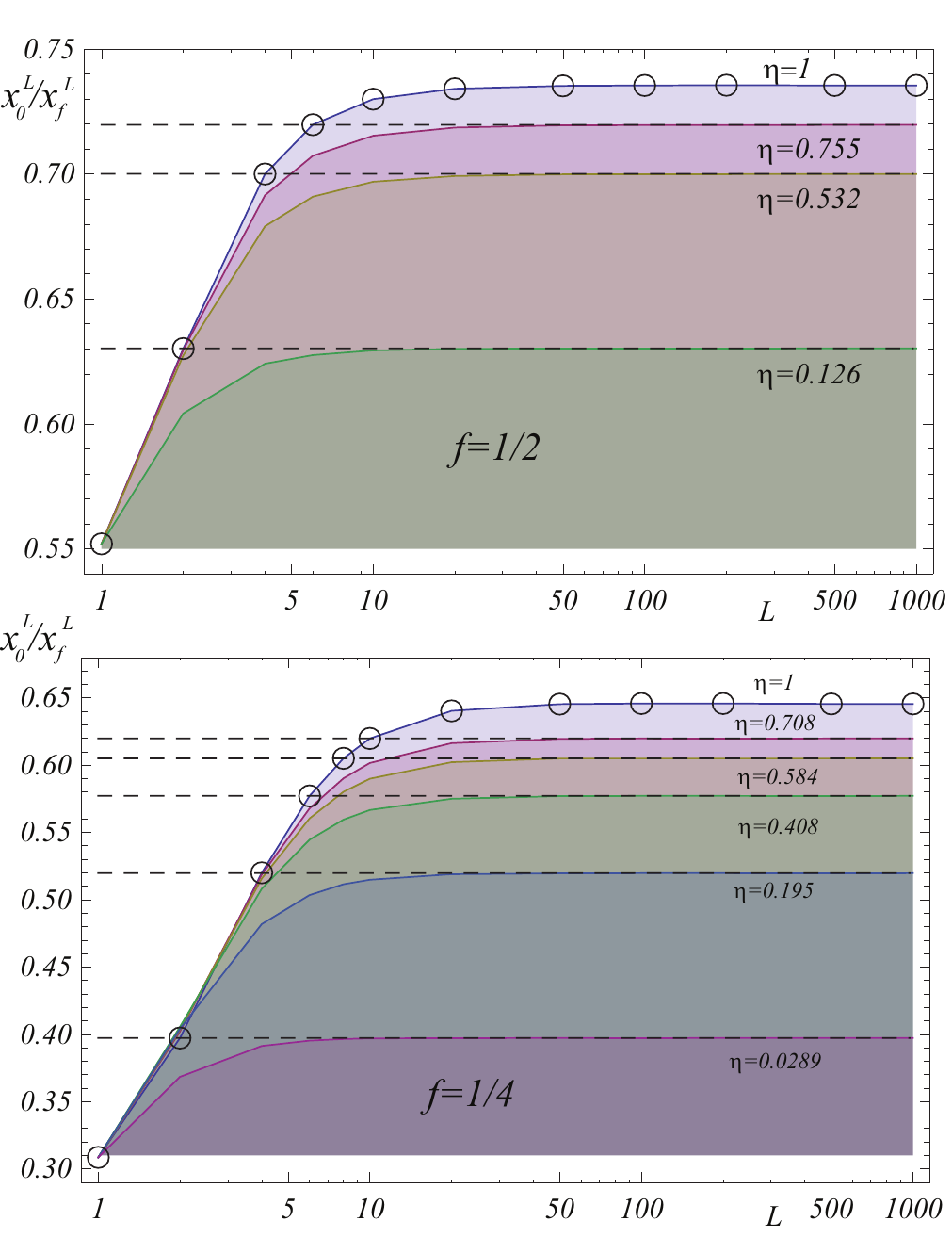}\caption{(Color online) The evolution of normalized critical ratio $x_{0}^{L}/x_{f}^{L}$
with number of layers for magnetic field $f=1/2$ and $f=1/4$ with
different values of the tunneling ratio $\eta=t_{\perp}/t_{\parallel}$.
Dashed lines show correspondence between systems with finite and infinite
$L$. }

\label{anisotropic} 
\end{figure}

\begin{figure}
\includegraphics[bb=0.69999999999999996cm 0.20000000000000001cm 10.1cm 13.199999999999999cm,clip,scale=0.85]{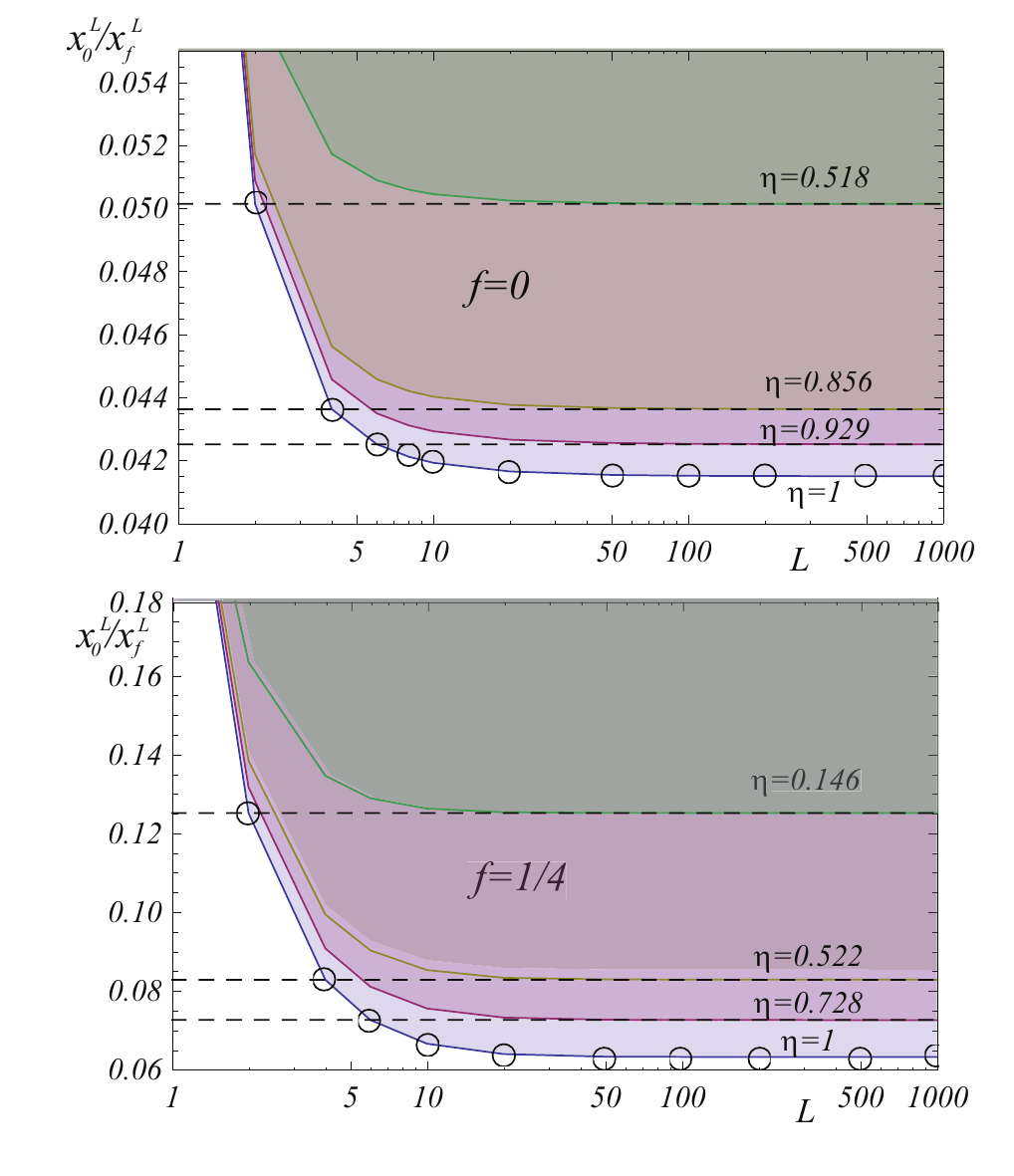}\caption{(Color online) The evolution of normalized critical parameter $x_{f}^{L}$
with number of layers without magnetic field and for $f=1/4$ with
different values of the tunneling ratio $\eta=t_{\perp}/t_{\parallel}$.
Dashed lines show correspondence between systems with finite and infinite
$L$. }

\label{anisotropic-1} 
\end{figure}
Although, conventional optical lattices typically possess a uniform
tunneling matrix elements, by changing lasers intensity along $z$
axis one can control the tunneling between adjacent layers. Sometimes
the tunneling rate between adjacent sites is just negligible on the
time scale of experiments \cite{kruger}. On the other hand, using
a mask in the novel holographic methods \cite{bakr} one can also
reproduce almost arbitrary potential. Such fully controllable optical
environment allows us to restrict the spatial (along $z$ axis) degree
of freedom of the particles and introduce $c$-axis anisotropy to
the system. In consequence, the tunneling between planes can be different
from that in the single layer $t_{\perp}\neq t_{\parallel}$ or even
completely suppressed. However, since some experimental setups can
be more convenient to use than others, once can try to establish a
correspondence between number of layers and anisotropy, i.e. find
values of $L$ and $\eta$ that result in the systems that share the
same properties in time-of-flight experiments, thus are interchangable.
The results are presented in Figs. \ref{anisotropic} and \ref{anisotropic-1}
for two setups: when the effect of the field is compared to the system
in zero field ($x_{0}^{L}/x_{f}^{L}$, see Fig. \ref{anisotropic}),
and when the bare critical value of $x_{f}^{L}$ is monitored (see,
Fig. \ref{anisotropic-1}). In both cases, we calculate $x_{0}^{L}/x_{f}^{L}$
or $x_{f}^{L}$ as a function of finite number of layers $L$ for
isotropic ($\eta=1$) system. Then, we determine the anisotropy ratio
$\eta$ for the infinite cubic system ($L\rightarrow\infty$), which
results in the same $x_{0}^{L}/x_{f}^{L}$ or $x_{f}^{L}$ values.
For example, from Fig. \ref{anisotropic} it is clear that 4-layer
(quasi 2D) isotropic system corresponds to 3D cubic system with anisotropy
ratio equal to $\eta=0.755$ ($f=1/2$, top plot). If the field value
$f=1/4$, a 3D system of similar anisotropy ($\eta=0.708$) shares
the properties with 10-layer isotropic one. Fig. \ref{anisotropic-1}
shows, that the $c$-axis anisotropy is less important when the magnetic
field is turned off (values of $\eta$ are closer to isotropic $\eta=1$
case).

\section{Conclusions}

We have calculated critical properties of the layered structure of
strongly interacting bosons under a synthetic magnetic field. Derived
phase diagram reveals a non-monotonic dependence of the ratio of tunneling
to on-site repulsive interaction on artificial magnetic field applied
to the system. By calculating analytically the density of states (cases
with $f=1/8$ and $f=3/8$ have not been presented in the literature)
for several values of the magnetic field we were able to accurately
predict the evolution of the system towards the Mott phase. The effect
is reduced when more layers are being added, i.e. during the two-
to quasi three-dimensional geometry crossover. Furthermore, we have
established a correspondence between anisotropic infinite (quasi three-dimensional)
and isotropic finite (slab geometry) systems that share exactly the
same critical values, which can be an important clue for choosing
experimental setups that are less demanding, but still leading to
the identical results. Finally, we have shown that the properties
of the ideal Bose gas in three-dimensional optical lattice can be
closely mimicked by finite (slab) systems, when the number of two-dimensional
layers is larger than ten or even less, when the layers are weakly
coupled.

\textbf{Acknowledgments}. T.A.Z. wants to acknowledge that the present
work is supported from scientific financial resources in the years
2009-2012 as a research grant.

\end{document}